\documentclass[10pt,conference]{IEEEtran}
\IEEEoverridecommandlockouts

\usepackage{cite}
\usepackage{booktabs}
\usepackage{multirow}
\usepackage{amsmath,amssymb,amsfonts}
\usepackage{graphicx}
\usepackage{textcomp}
\usepackage{url}
\usepackage{xspace}
\usepackage[table]{xcolor}
\usepackage{float}
\usepackage{caption}
\usepackage{listings}
\usepackage{stfloats}
\usepackage{enumitem}
\usepackage[most]{tcolorbox}
\definecolor{keycolor}{RGB}{0,0,180}
\definecolor{valuecolor}{RGB}{220,35,35}
\definecolor{gptbestbg}{RGB}{224,247,224}
\definecolor{dsbestbg}{RGB}{255,235,214}
\definecolor{goldrowbg}{RGB}{232,232,232}
\newcommand{\gptbest}[1]{\begingroup\setlength{\fboxsep}{1pt}\colorbox{gptbestbg}{#1}\endgroup}
\newcommand{\dsbest}[1]{\begingroup\setlength{\fboxsep}{1pt}\colorbox{dsbestbg}{#1}\endgroup}

\newcommand{\gptlegend}{\begingroup\setlength{\fboxsep}{1pt}\colorbox{gptbestbg}{GPT-5 mini}\endgroup}
\newcommand{\dslegend}{\begingroup\setlength{\fboxsep}{1pt}\colorbox{dsbestbg}{DeepSeek V3.2}\endgroup}
\newcommand{\abldown}[1]{{\textcolor{green!50!black}{(#1)}}}
\newcommand{\ablup}[1]{{\textcolor{orange!85!black}{(#1)}}}
\newcommand{\ablflat}[1]{{\textcolor{green!50!black}{(#1)}}}
\lstdefinestyle{artifactjson}{
  basicstyle=\ttfamily\scriptsize,
  columns=fullflexible,
  breaklines=true,
  breakatwhitespace=false,
  showstringspaces=false,
  keepspaces=true,
  frame=none,
  xleftmargin=0pt,
  aboveskip=3pt,
  belowskip=3pt,
  escapeinside={(*@}{@*)}
}
\newtcolorbox{promptbox}[1]{
  enhanced,
  breakable,
  colback=blue!4!white,
  colframe=blue!70!black,
  colbacktitle=blue!85!black,
  coltitle=white,
  fonttitle=\bfseries,
  fontupper=\footnotesize,
  title={#1},
  boxrule=0.9pt,
  arc=2pt,
  left=6pt,
  right=6pt,
  top=6pt,
  bottom=6pt,
  before skip=6pt,
  after skip=8pt
}
\newtcolorbox{findingbox}[1]{
  enhanced,
  breakable,
  colback=orange!6!white,
  colframe=orange!70!black,
  colbacktitle=orange!85!black,
  coltitle=white,
  fonttitle=\bfseries,
  title={#1},
  boxrule=0.9pt,
  arc=2pt,
  left=6pt,
  right=6pt,
  top=6pt,
  bottom=6pt,
  before skip=6pt,
  after skip=8pt
}
\newcommand{\silin}[1]{{\textcolor{red}{[Silin: #1]}}}
\newcommand{\approach}{{Repo0}\xspace}
\newcommand{\supmatfootnote}{\footnote{\url{https://github.com/cslsolow/Repo0/blob/main/supplementary.pdf}}}
\def\BibTeX{{\rm B\kern-.05em{\sc i\kern-.025em b}\kern-.08em
    T\kern-.1667em\lower.7ex\hbox{E}\kern-.125emX}}

\newcommand{\gu}[1]{\textcolor{blue}{[Gu: #1]}}

\begin{document}
\bstctlcite{IEEEexample:BSTcontrol}

\title{\approach: Design-Driven Zero-to-All Code Generation}
\author{
\IEEEauthorblockN{Silin Chen\textsuperscript{1,*}, Haoyi Teng\textsuperscript{1,*}, Xiaodong Gu\textsuperscript{1,\ensuremath{\dagger}}, Yuling Shi\textsuperscript{1}, Jiale Huang\textsuperscript{1}, Yongpan Wang\textsuperscript{1},\\
Hongyu Zhang\textsuperscript{2}, Haibing Guan\textsuperscript{1}}
\IEEEauthorblockA{\textsuperscript{1}Shanghai Jiao Tong University\\
\textsuperscript{2}Chongqing University}
}

\maketitle
\begingroup
\renewcommand{\thefootnote}{\fnsymbol{footnote}}
\footnotetext[1]{Silin Chen and Haoyi Teng contributed equally.}
\footnotetext[2]{Xiaodong Gu is the corresponding author.}
\endgroup
\begingroup
\renewcommand{\thefootnote}{}
\footnotetext{Emails: \url{cslsolow@gmail.com}, \url{chunemeng@outlook.com}, \url{xiaodong.gu@sjtu.edu.cn}, \url{yuling.shi@sjtu.edu.cn}, \url{unicornshjl@gmail.com}, \url{frankile@sjtu.edu.cn}, \url{hbguan@sjtu.edu.cn}, and \url{hongyujohn@gmail.com}.}
\endgroup

\begin{abstract}
Large language model agents have made substantial progress in code generation, yet most existing systems assume a predefined repository architecture. This assumption does not hold in zero-to-all code generation, where an agent must construct an entire software project directly from natural-language requirements while maintaining a modular repository architecture throughout development.
We present \approach, a continuous structural evolution framework for zero-to-all code generation. \approach maintains an explicit architectural state instantiated as a Dual-Directed-Acyclic-Graph (Dual-DAG), consisting of a requirement-level DAG, a component-level DAG, and their alignment relation. Starting from natural-language requirements, it iteratively evolves component boundaries through structural actions guided by modularity metrics until structural convergence, after which the converged architecture guides test-driven development code generation.
We evaluate \approach on six real-world repositories from RepoCraft using GPT-5 mini and DeepSeek V3.2. \approach achieves the highest Functionality Coverage and Pass Rate across all settings. Compared with RPG, the strongest repository-planning baseline, \approach improves Functionality Coverage by up to 20.08 percentage points and Pass Rate by up to 29.74 percentage points. Ablation and structural-evolution analyses further demonstrate the importance of the Dual-DAG architectural state, modularity-guided structural evolution, and explicit structural convergence\footnote{Our code and data are available at \url{https://github.com/cslsolow/Repo0}}.
\end{abstract}


\section{Introduction}


Code generation has seen remarkable progress with the advent of Large Language Models (LLMs), leading to highly capable agents that can synthesize complex code snippets and resolve repository-level issues~\cite{jimenez2024swebench,yang2024sweagent,bairi2023codeplan,wang2024openhands,peng2025swe,chen2025swe,ma2026llmagentscoderepositories,li2025aligning,shi2026swebenchpromaxbenchmarkingagents}. However, current methods predominantly operate under a critical assumption: they assume the code has already been architecturally designed. In these structured repository completion settings, agents operate on top of predefined or partially specified repository architectures, where key design decisions, such as component boundaries, package organization, and dependency graphs, are largely given~\cite{zhao2024commit0}. By focusing on the coding phase, these approaches overlook the crucial role of software architectural design.

\begin{figure}[!t]
  \centering
  \includegraphics[width=\columnwidth]{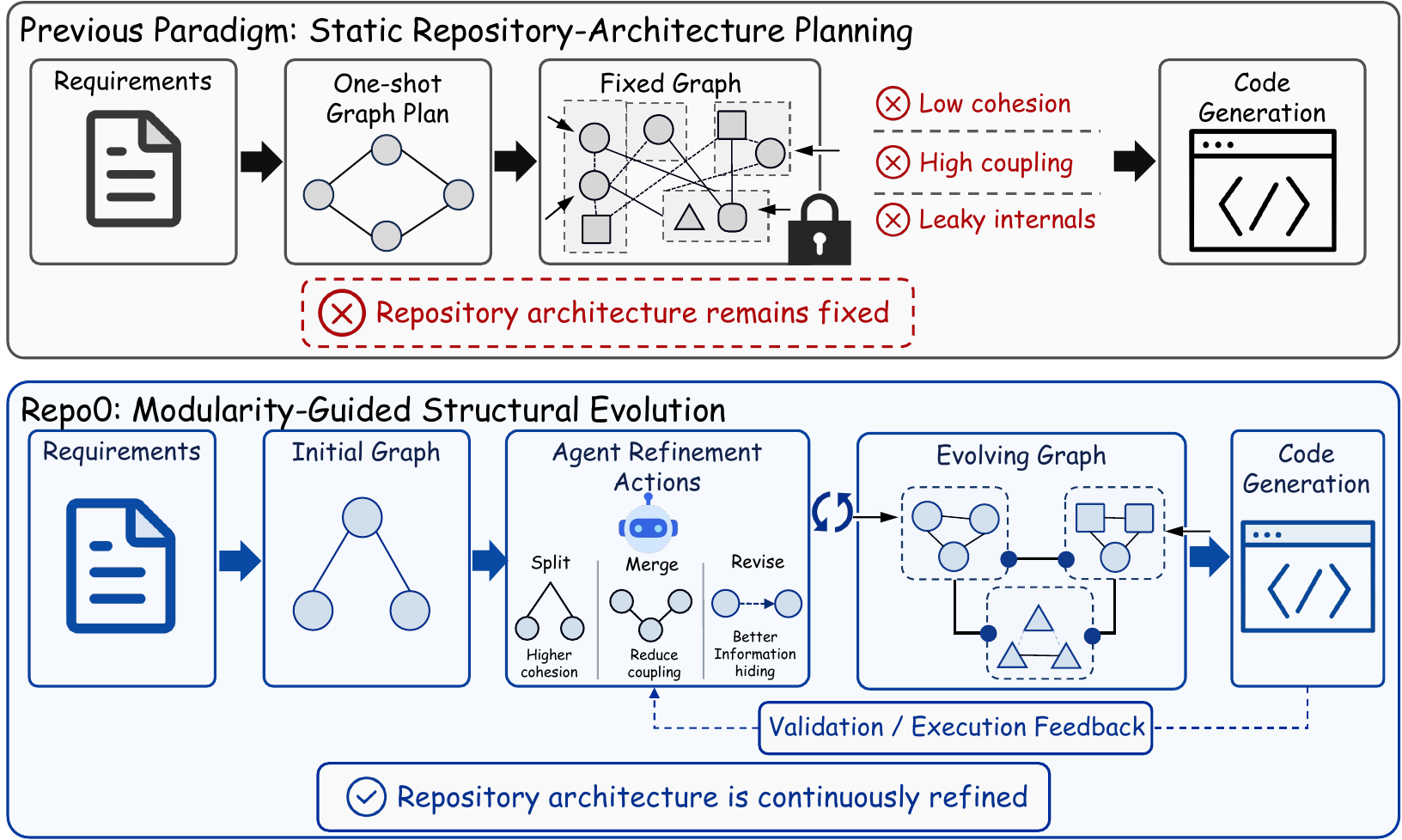}
  \caption{Previous methods treat the graph as a fixed planning artifact, whereas \approach continuously evolves repository architecture during generation.}
  \label{fig:intro-motivation}
  \vspace{-12pt}
\end{figure}

In this paper, we focus on a more challenging and realistic problem: \textit{zero-to-all code generation}. We seek to construct software projects without pre-existing repository architecture~\cite{luo2025rpg,peng2026repogenesis}. 
This problem is difficult because the agent must jointly infer both the software's functionality and its architecture. A well-designed software project must adhere to core software engineering principles such as high cohesion and low coupling~\cite{yourdon1979structured}. 
When a coding agent attempts zero-to-all code generation without explicit architectural guidance, it could violate these modularity goals, resulting in inconsistent component boundaries, fragile dependency structures, and poor cross-file coordination. This suggests that the core bottleneck in zero-to-all code generation is not merely synthesizing code, but establishing and maintaining robust repository modularity throughout the development process.

Recent approaches have begun to explore this problem~\cite{zhao2024commit0,ding2025nl2repo,luo2025rpg,peng2026repogenesis,lu2026projdevbench}. Approaches like NL2Repo-Bench~\cite{ding2025nl2repo} bypass the design stage by providing a golden repository architecture as part of the input. Multi-agent systems attempt to coordinate planning and implementation through specialized roles~\cite{qian2024chatdev,hong2024metagpt,ishibashi2024self}.  Recent graph-based approaches introduce explicit repository planning~\cite{luo2025rpg}. Despite these advances, these approaches suffer from a critical bottleneck: they treat software design as a \textit{static} artifact. They assume that a perfect architectural blueprint can be generated during a single, initial planning stage and then rigidly executed. 
In real-world software development, however, project architecture is rarely ready during initial planning. As development progresses, the generated code often reveals whether planned components are truly cohesive or if dependency edges are overly entangled. Emerging complexities, duplicated functionalities, and validation feedback frequently necessitate splitting low-cohesion components or merging redundant ones. Consequently, zero-to-all code generation cannot be treated as a one-shot planning problem; it must be a continuous structural evolution process (as illustrated in Figure~\ref{fig:intro-motivation}).

\begin{figure*}[!t]
  \centering
  \includegraphics[width=0.92\textwidth]{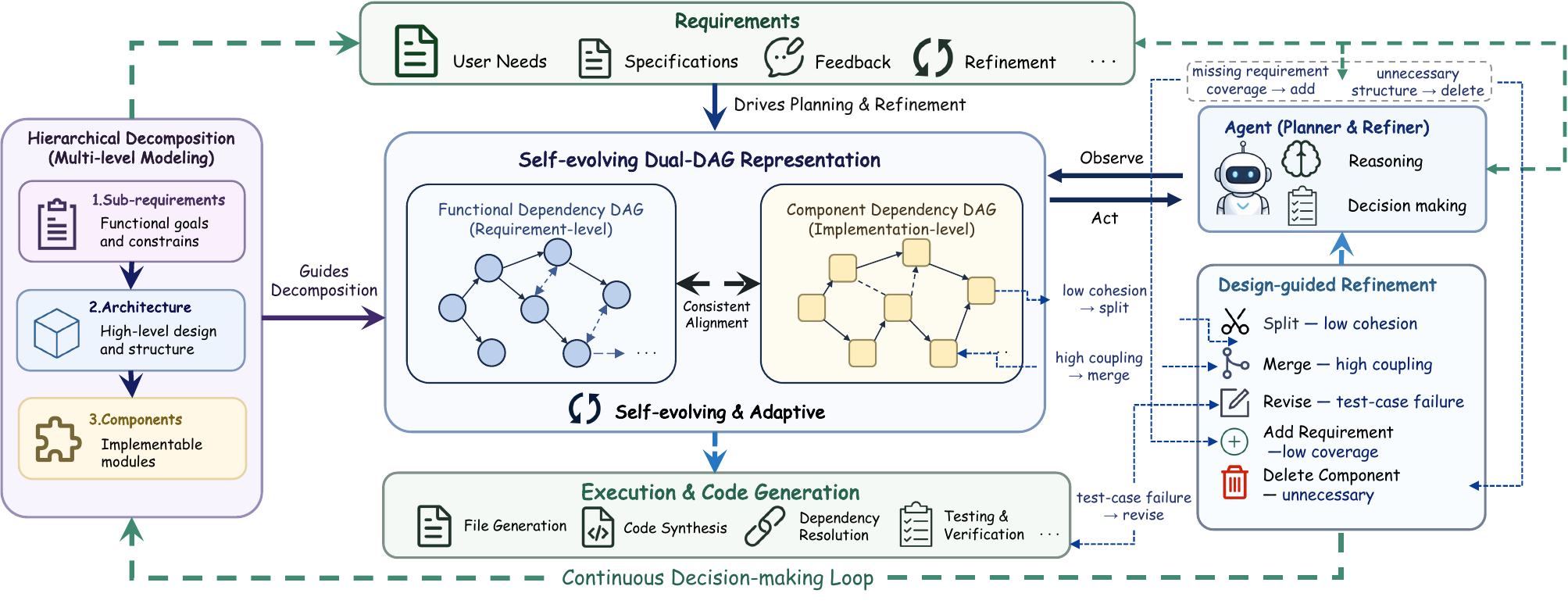}
  \caption{Overview of the continuous decision-driven structural evolution framework.}
  \label{fig:method-overview}
\end{figure*}

To address this limitation, we propose \approach, a design-driven structural evolution framework for zero-to-all code generation. Rather than committing to a rigid initial blueprint, \approach explicitly models repository generation as a continuous structural evolution problem. Specifically, we maintain an architectural state represented as a Dual-Directed-Acyclic-Graph (Dual-DAG). The state separates requirement-level functional relationships from implementation-level components and dependencies, allowing functional coordination and implementation dependencies to evolve without being conflated in a single static planning graph. This Dual-DAG preserves traceability from the initial user prompt down to the code, while keeping the project architecture open to subsequent structural updates. 
Starting from a blank state with only natural language requirements, \approach executes a full-lifecycle software development process. It first extracts, normalizes, and decomposes the requirements to construct a requirement-level DAG. Then, it incrementally evolves the component-level DAG before code generation and uses the converged architecture to guide repository generation. Through explicit structural actions (i.e., \texttt{add}, \texttt{split}, \texttt{merge}, \texttt{revise}, 
and \texttt{save}), \approach dynamically adjusts component boundaries based on modularity metrics. After structural convergence, validation feedback drives localized repair during code generation. This evolution process is guided by modularity metrics and continues until structural convergence is reached. By treating design and implementation as deeply intertwined, \approach represents a paradigm shift in how autonomous agents construct complex software systems from scratch.

We evaluate \approach on RepoCraft~\cite{luo2025rpg,luo2026closing}, a repository-generation benchmark consisting of six real-world Python repositories, under both GPT-5 mini and DeepSeek V3.2. Results show that \approach consistently improves repository-generation quality over baselines. Across all six repositories and both backbone models, \approach achieves the highest Functionality Coverage and Pass Rate in all settings. Compared with RPG \cite{luo2025rpg}, the strongest repository-planning baseline, \approach improves Functionality Coverage by 4.55--20.08 percentage points and Pass Rate by 7.61--29.74 percentage points. Ablation results show that each major design component contributes to end-to-end performance, while the structural-evolution analysis further indicates that metrics-guided convergence is crucial: cohesion- and coupling-guided evolution improves repository quality, whereas unconstrained LLM-decided structural actions tend to over-decompose the repository architecture and degrade downstream correctness.

Overall, this paper makes three contributions:
\begin{enumerate}

\item We identify a fundamental limitation of existing repository generation methods: they treat repository architecture as a static artifact determined before implementation. We argue that repository modularity is only partially observable from requirements and emerges throughout coding. We formulate repository generation inherently as a continuous structural evolution problem rather than a one-shot planning problem.

\item We reformulate zero-to-all repository generation as a modularity-driven structural evolution process. Instead of committing to a fixed repository plan, repository architecture is continuously evolved through explicit structural actions guided by modularity metrics and requirement-coverage checks until structural convergence is reached. After convergence, validation feedback drives localized repair during code generation.

\item We instantiate this formulation in \approach, a design-driven repository-generation framework built upon a Dual-DAG representation that separates requirement-level functionalities from implementation-level components and dependencies while maintaining end-to-end traceability across requirements, architecture, and code.

\end{enumerate}

\section{Methodology}
\label{sec:methodology}

\subsection{Problem Formulation}

We frame zero-to-all repository generation as a problem of continuous structural evolution. Unlike prior methods that first design a repository architecture and then treat it as a fixed blueprint for code generation, \approach treats repository design as an adaptive process whose structure is continuously revised as new architectural evidence emerges during requirement decomposition, component construction, code generation, and validation. The central objective is not only to capture complex long-context relationships across a repository for correct code generation, but also to establish and improve repository modularity throughout the development lifecycle.

Figure~\ref{fig:method-overview} illustrates the overall framework. To support continuous structural evolution, \approach maintains a persistent architectural state $S_t = (G_t^R, G_t^C, \mathcal{A}_t)$ for each time step $t$.

$\mathbf{G_t^R = (V_t^R, E_t^R)}$ is a requirement-level DAG where \(V_t^R\) represents a requirement unit, including both high-level requirements and their sub-requirements. Edges in this DAG encode requirement-level functional coordination relations instead of implementation dependencies: an edge $(u,v) \in E_t^R$ indicates that two requirements are logically used together and should therefore be considered jointly when aligning their behavior, inputs, and outputs. 

$\mathbf{G_t^C = (V_t^C, E_t^C)}$ is the component-level DAG, where each \(v\in V_t^C\) represents a component, such as a module, object, service layer, parser, or adapter, that can later be materialized as concrete source code. A component is the basic design unit that represents a bounded implementation responsibility (e.g., a reusable ``Unified Modeling API'') that will later be materialized into concrete repository artifacts. \(E_t^C\) encodes implementation dependencies such as inheritance, reuse, or containment. 
    
$\mathbf{\mathcal{A}_t \subseteq V_t^R \times V_t^C}$ denotes the relations between requirements and components. A pair $(q, c) \in \mathcal{A}_t$ indicates that component $c$ realizes all or part of requirement $q$. The relation is many-to-many: a requirement may require several components, and a reusable component may support multiple requirements. The alignment relation preserves traceability from requirements to components throughout structural evolution.

 \(S_t\) is dynamic over time \(t\): while the requirement-level DAG $G_t^R$ generally stabilizes early, the component-level DAG $G_t^C$ and the alignment relation $\mathcal{A}_t$ are continuously updated until modularity is optimized.

\subsection{Phase I: Requirement Decomposition and Initial Architecture}

The evolution process begins by translating a natural-language requirement document $D$ into the initial architectural state $S_0$. The requirement documents are usually uneven in granularity: some phrases describe broad project goals, while others describe constraints, behaviors, or usage scenarios. Directly mapping such descriptions to design modules often leaves important requirements under-specified and encourages brittle repository architectures.

\begin{figure}[!t]
  \centering
  \includegraphics[width=\columnwidth]{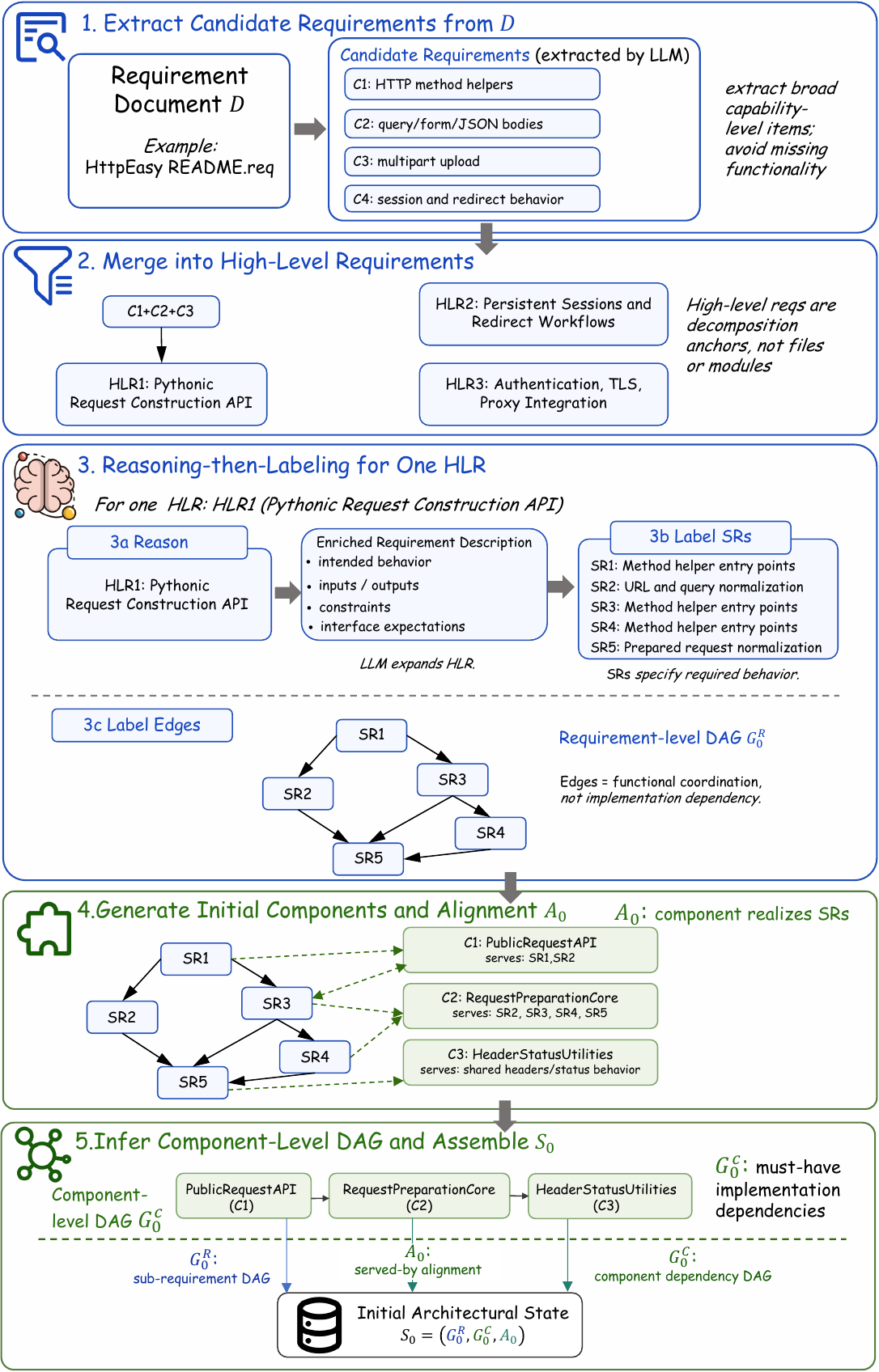}
  \caption{Illustrative construction of the initial architectural state.}
  \label{fig:phase1-example}
  \vspace{-8pt}
\end{figure}

To resolve this mismatch, \approach first extracts candidate requirement items from $D$. Specifically, an LLM is prompted to identify capability-level requirements under the following regularizations: each item should describe a distinct repository-level functionality or system constraint, include the relevant operations and sub-features in its description, and avoid splitting low-level operations into separate top-level items. \approach then applies a requirement-merge step that consolidates redundant or subsumed items while keeping merely related requirements separate. The remaining merged items are treated as high-level requirements. They serve as top-level decomposition anchors rather than direct file or module specifications. 

Next, each high-level requirement is decomposed into sub-requirements. A \emph{sub-requirement} refers to a requirement-side functional unit that clarifies \textit{what} functionality must be provided without determining packages, files, classes, or component dependencies. Inspired by Atom of Thoughts~\cite{teng2026atom}, \approach decomposes requirements through a three-stage reasoning-then-labeling process: First, an LLM is prompted to produce as comprehensive a requirement description as possible, elaborating the intended behavior, inputs and outputs, functional constraints, interface expectations, error cases, and ambiguous scope. Based on this enriched requirement description, \approach then prompts the LLM to identify sub-requirements and finally label their logical dependency relations. A directed edge from sub-requirement $u$ to sub-requirement $v$ is added when $v$ logically follows $u$, such as when $v$ relies on the behavior, output, data contract, or interface assumption established by $u$. The resulting nodes and labeled edges form the local requirement-level DAG for that high-level requirement.
Together, the high-level requirements and their decomposed sub-requirements form the initial requirement-level DAG, $G_0^R$.

Once $G_0^R$ is constructed, the system derives the initial components to populate $G_0^C$ and establishes the initial alignment relation $\mathcal{A}_0$. Concretely, for each high-level requirement, \approach collects its sub-requirements from $G_0^R$ and prompts the LLM to translate them into a set of bounded components. Each proposed component contains a name, a responsibility description, and an explicit list of served sub-requirements. The proposed components form the initial component nodes $V_0^C$, and each served-sub-requirement entry creates an alignment pair $(q,c)\in\mathcal{A}_0$. Finally, \approach prompts the LLM to infer the implementation dependencies among the proposed components. Requirement-level coordination edges in $G_0^R$ are provided only as soft evidence rather than being directly copied into $G_0^C$. The inferred dependencies initialize $E_0^C$. This initialization separates requirement-side artifacts from implementation artifacts.

Figure~\ref{fig:phase1-example} illustrates how Phase I builds the initial architecture for a zero-to-all \textit{HttpEasy} repository. Starting from $D$, \approach first extracts candidate requirement items and merges overlapping or subsumed items into high-level requirements. In this example, request construction, session persistence, response handling, and authentication are consolidated into two high-level requirements: ``HTTP request and response core'' and ``session and authentication management''.
For a selected high-level requirement, the LLM then produces an enriched semantic specification that elaborates parameters, headers, body encoding, timeout behavior, error handling, and response objects. Based on this specification, it identifies sub-requirements including constructing a request object, sending the request through a transport adapter, and parsing the returned response, and organizes them into a functional coordination chain from request construction to transport execution and response parsing, thereby instantiating $G_0^R$.
These sub-requirements are then grounded into implementation components such as \texttt{RequestBuilder}, \texttt{TransportAdapter}, and \texttt{ResponseParser}. For instance, \texttt{RequestBuilder} serves the sub-requirement of constructing a request object, which establishes a corresponding entry in the alignment relation $\mathcal{A}_0$. Finally, \approach infers must-have implementation dependencies among components---for example, \texttt{SessionClient} depends on \texttt{RequestBuilder} and \texttt{TransportAdapter}---to construct $G_0^C$ and yield the initial architectural state $S_0=(G_0^R,G_0^C,\mathcal{A}_0)$.

\subsection{Phase II: Modularity-Guided Structural Evolution Loop}

The initial architectural state provides an initial assignment of requirement nodes to components, but may still be overly broad, fragmented, or redundant. \approach therefore evolves $G_t^C$ and \(\mathcal{A}_t\) through an evolution loop. During this phase, $G_t^R$ is fixed to preserve the functional scope, while $G_t^C$ and $\mathcal{A}_t$ are updated to improve repository modularity.
The evolution is realized through four component-level structural actions:
\begin{itemize}
    \item \texttt{split}: For a diffuse component $c$, \approach extracts the induced subgraph of $G_t^R$ over its served sub-requirements $S(c)$ and uses graph partitioning~\cite{bichot2013graph}, viewed as a minimum-cut objective~\cite{wagner1993between}, to identify candidate sub-requirement groups. The partition evidence is provided to the LLM, which rewrites $c$ into a set of narrower components, redistributes the corresponding alignment pairs in $\mathcal{A}_t$, and reconnects incident dependencies in $G_t^C$.
    \item \texttt{merge}: Consolidates two components into a single component by combining their responsibility descriptions, merging their alignment pairs in $\mathcal{A}_t$, and redirecting incident dependencies in $G_t^C$ to the merged component.
    \item \texttt{revise}: A boundary-preserving action that rewrites a component's responsibility description, interface assumptions, implementation notes, or alignment entries.
    \item \texttt{save}: Marks a component as structurally stable and carries it forward unchanged in the current round unless neighboring structural updates make it eligible again.
\end{itemize}

These structural evolution actions are triggered by two modularity metrics and two auxiliary criteria: 

\paragraph{Responsibility}
Responsibility identifies the sub-requirements assigned to a component. For a component \(c\), we define
\begin{equation}
RS(c)=\{q \in V_t^R \mid (q,c)\in \mathcal{A}_t\}.
\end{equation}
The size \(|RS(c)|\) measures how many requirement-side units that \(c\) is responsible for.

\paragraph{Cohesion} 
Cohesion identifies components that group diffuse, unrelated responsibilities~\cite{stevens1974structured}. 
Let $E_{\mathrm{in}}(c)$ denote the number of edges in $G_t^R$ whose endpoints both lie strictly within $RS(c)$. We define cohesion as the density of realized requirement-level functional coordination relations among the requirements grouped into the same component:
\begin{equation}
\mathrm{cohesion}(c)=
\begin{cases}
1, & |RS(c)| \le 1\\
\dfrac{E_{\mathrm{in}}(c)}{|RS(c)|(|RS(c)|-1)/2}, & |RS(c)| > 1
\end{cases}
\end{equation}
Low cohesion indicates a structurally diffuse component. 

A \texttt{split} action is triggered when a component's cohesion falls below a threshold $\gamma_{\mathrm{split}}$ and the size of $RS(c)$ exceeds a threshold $\tau_{\mathrm{split}}^{(t)}$. 
Here, $\tau_{\mathrm{split}}^{(t)}$ is used to prevent the loop from over-splitting.


\paragraph{Coupling}
Coupling identifies pairs of components that realize highly overlapping sets of sub-requirements~\cite{stevens1974structured}. Let $RS_A$ and $RS_B$ be the sub-requirement sets for which components $A$ and $B$ are responsible, respectively. Coupling is defined as the Jaccard Similarity~\cite{real1996probabilistic} between these sets:
\begin{equation}
\mathrm{coupling}(A,B)=\frac{|RS_A\cap RS_B|}{|RS_A\cup RS_B|}
\end{equation}

\paragraph{Connectivity}
Connectivity measures the edges in $G_t^R$ bridging $RS_A$ and $RS_B$. 
A component pair is considered a \texttt{merge} candidate when their coupling exceeds a threshold \(\theta_\text{merge}\) and connectivity is greater than 1. 
These metrics alone may neglect duplicated component responsibilities from valid architectural coupling, such as layered, adapter, or upstream-downstream relationships. Therefore, we ask an LLM to validate the candidates and determine the final \texttt{merge} action.

The structural evolution loop proceeds iteratively. In each round, \approach recomputes the modularity metrics and auxiliary criteria, then applies eligible structural actions to $G_t^C$. 

The iteration terminates when a complete evolution round yields no eligible \texttt{split} or \texttt{merge} actions. At this point, structural convergence is reached with respect to component boundaries. The system then performs a final semantic alignment pass, where the LLM examines each component together with its aligned sub-requirements and neighboring components in the component-level DAG. If inconsistencies are identified between component responsibilities, requirement coverage, or interface assumptions, \texttt{revise} actions are applied.

\subsection{Phase III: Code Generation}
Once structural convergence is reached, the repository architecture is fixed, and \approach enters a code generation phase.
The system first converts the converged $G_t^C$ into a concrete generation plan: it derives package assignments, planned file paths, exported symbols, and dependency-aware generation order from $G_t^C$. For each component, \approach constructs a generation context from three sources: the component's responsibility description, its aligned requirement nodes recorded in $\mathcal{A}_t$, and upstream components that must be available before the current component is implemented. This context specifies the component's target behavior, placement, public symbols, and reusable dependencies.

Code is generated incrementally under a Test-driven development (TDD) workflow~\cite{chang2026test,beck2003test}. For each component, \approach first generates an importable skeleton that fixes the expected public API, then synthesizes tests from the aligned requirement nodes and interface assumptions, and finally fills in the implementation to satisfy those tests. After each generation step, the system runs validation checks, including import checks, interface checks, and pytest-based execution. Failures such as missing symbols, incompatible call signatures, or unmet behavioral expectations trigger localized repair patches to the affected implementation, tests, or package initialization files. When validation feedback reveals that the component description or interface assumptions are inaccurate, \approach can apply \texttt{revise} before the next TDD workflow. 
\section{Experimental Setup}
\subsection{Research Questions}
We study three research questions:

\noindent\textbf{RQ1 (The Overall Effectiveness).} How effective is \approach at zero-to-all code generation?

\noindent\textbf{RQ2 (Ablation Study).} How does each core design component of \approach contribute to zero-to-all code generation?

\noindent\textbf{RQ3 (Structural Evolution Analysis).} Do the proposed modularity metrics improve structural convergence compared with LLM-decided structural actions?

\subsection{Datasets}

\begin{table}[t]
\caption{Overview of the six repositories and their paraphrased
counterparts (\textit{Para. Name}) in RepoCraft. \#Files denotes the total
source files, LOC the effective lines of code, and Task Counts the evaluation
tasks.}
\label{tab:repocraft-datasets}
\centering
\footnotesize
\begin{tabular*}{\columnwidth}{@{\extracolsep{\fill}}llrrr@{}}
\toprule
Real Repo & Para. Name & \#Files & LOC & Task Counts \\
\midrule
scikit-learn & MLKit-Py & 185 & 65,972 & 236 \\
pandas & TableKit & 217 & 106,447 & 175 \\
sympy & SymbolicMath & 699 & 218,924 & 192 \\
statsmodels & StatModeler & 271 & 83,325 & 234 \\
requests & HttpEasy & 17 & 2,793 & 50 \\
django & PyWebEngine & 681 & 109,457 & 165 \\
\bottomrule
\end{tabular*}
\end{table}

We evaluate \approach on RepoCraft~\cite{luo2025rpg}, a repository-generation benchmark consisting of six real-world Python repositories. Following RPG~\cite{luo2025rpg}, repositories are exposed through paraphrased counterpart names to reduce potential pretraining leakage.

To better align the benchmark with the specific zero-to-all code generation task, namely, generating a repository from high-level requirements, we further ask two software engineers to revise the original task descriptions~\cite{luo2026closing}. The revised descriptions remove repository-structure cues, implementation-specific architectural hints, and function-level interface details while preserving the original functional requirements.

The benchmark includes \textit{scikit-learn}, \textit{pandas}, \textit{sympy}, \textit{statsmodels}, \textit{requests}, and \textit{django}, exposed as \textit{MLKit-Py}, \textit{TableKit}, \textit{SymbolicMath}, \textit{StatModeler}, \textit{HttpEasy}, and \textit{PyWebEngine}, respectively. As summarized in Table~\ref{tab:repocraft-datasets}, these repositories span diverse software domains and scales, providing a challenging benchmark for evaluating repository generation across both compact and dependency-intensive systems.

\subsection{Baseline Methods}

We compare \approach with three representative baselines covering direct coding agents, staged repository generation, and graph-based repository planning. We additionally report the original ground-truth repository for each benchmark as a reference.

\begin{itemize}[leftmargin=*, itemsep=3pt, topsep=2pt]
  \item \textbf{mini-SWE-agent} is a lightweight software engineering agent that solves coding tasks through iterative code editing and command execution~\cite{yang2024sweagent,minisweagent2026}.

  \item \textbf{Paper2Code} is a staged multi-agent repository generation framework that decomposes generation into planning, analysis, and implementation phases~\cite{seo2025paper2code}.

  \item \textbf{RPG} is the state-of-the-art graph-based repository generation method, which constructs a static Repository Planning Graph to guide code generation~\cite{luo2025rpg}.

  \item \textbf{Gold Project} is the original repository corresponding to each benchmark task and serves as a reference for validating the evaluation pipeline.
\end{itemize}

\subsection{Metrics}

Following RPG~\cite{luo2025rpg}, we adopt its evaluation pipeline and metric definitions.

\textbf{Functionality Coverage} measures the fraction of reference functional categories matched by at least one generated functionality description. 

\textbf{Functionality Novelty} measures the fraction of generated functionalities that are not matched to any reference category under the same matching procedure. 

\textbf{Functionality Accuracy} is evaluated using two task-level metrics: \textit{Pass Rate}, the fraction of benchmark tasks whose adapted ground-truth tests pass on the generated repository, and \textit{Voting Rate}, the fraction of tasks for which majority-vote semantic evaluation identifies a matched functional interface.

To mitigate evaluator bias, we adopt cross-model evaluation for both semantic voting and test-case rewriting. Specifically, DeepSeek V3.2 evaluates the repositories and rewritten test cases generated by GPT-5 mini, while GPT-5 mini evaluates those generated by DeepSeek V3.2.


\begin{table*}[!t]
\caption{RQ1 main results on three RepoCraft repositories. For each repository, we report Functionality Coverage (Cov.), Functionality Novelty (Nov.), and Pass./Vot., which combines Pass Rate and Voting Rate. \gptlegend\ marks the highest value among methods under GPT-5 mini, and \dslegend\ marks the highest value among methods under DeepSeek V3.2.}
\label{tab:rq1-main}
\centering
\scriptsize
\setlength{\tabcolsep}{2.5pt}
\resizebox{\textwidth}{!}{%
\begin{tabular}{llccccccccc}
\toprule
\multirow{2}{*}{Model} & \multirow{2}{*}{Method} & \multicolumn{3}{c}{requests} & \multicolumn{3}{c}{statsmodels} & \multicolumn{3}{c}{django} \\
\cmidrule(lr){3-5} \cmidrule(lr){6-8} \cmidrule(lr){9-11}
 &  & Cov. (\%) & Nov. (\%) & Pass./Vot. (\%) & Cov. (\%) & Nov. (\%) & Pass./Vot. (\%) & Cov. (\%) & Nov. (\%) & Pass./Vot. (\%) \\
 \midrule
 \multirow{5}{*}{GPT-5 mini} & mini-SWE-agent & 68.18 & 2.63 & 4.11 / 27.40 & 18.18 & 9.09 & 0.00 / 31.86 & 47.92 & 6.96 & 37.04 / 44.44 \\
 & Paper2Code & 95.50 & 7.20 & 24.66 / 24.66 & 44.32 & \gptbest{24.13} & 4.42 / 30.09 & 66.67 & \gptbest{15.30} & 30.04 / 78.60 \\
 & RPG & 90.91 & 13.70 & 31.51 / \gptbest{95.89} & 70.40 & 13.80 & 77.90 / 92.00 & 60.42 & 11.58 & 47.33 / 74.07 \\
& \approach & \gptbest{100.00} & \gptbest{18.20} & \gptbest{50.98} / \gptbest{100.00} & \gptbest{80.68} & 11.48 & \gptbest{85.51} / \gptbest{98.65} & \gptbest{80.50} & 13.59 & \gptbest{74.36} / \gptbest{97.12} \\
\midrule
\multirow{5}{*}{DeepSeek V3.2} & mini-SWE-agent & 86.36 & 15.34 & 21.92 / 47.95 & 59.09 & \dsbest{23.39} & 2.65 / 53.98 & 33.33 & 9.38 & 10.70 / 47.33 \\
 & Paper2Code & 90.91 & 11.80 & 4.11 / 56.16 & 14.77 & 5.00 & 49.56 / 61.95 & 62.50 & \dsbest{43.46} & 7.82 / 53.50 \\
 & RPG & 95.45 & 9.23 & 61.64 / 90.41 & 64.70 & 13.70 & 39.29 / 73.57 & 68.75 & 26.50 & 46.50 / 69.55 \\
 & \approach & \dsbest{100.00} & \dsbest{24.77} & \dsbest{78.08} / \dsbest{100.00} & \dsbest{78.41} & 14.10 & \dsbest{69.03} / \dsbest{86.46} & \dsbest{79.17} & 14.29 & \dsbest{74.07} / \dsbest{93.83} \\
\midrule
\rowcolor{black!8}
\multirow{1}{*}{Human Developer} & Gold Project & \multicolumn{1}{c}{100.00} & \multicolumn{1}{c}{--} & \multicolumn{1}{c}{94.12 / 100.00} & \multicolumn{1}{c}{100} & \multicolumn{1}{c}{--} & \multicolumn{1}{c}{94.15 / 100.00} & \multicolumn{1}{c}{100.00} & \multicolumn{1}{c}{--} & \multicolumn{1}{c}{96.34 / 100.00} \\
\bottomrule
\end{tabular}%
}
\end{table*}

\subsection{Implementation Details}

We use two backbone models in our experiments: the open-source DeepSeek V3.2~\cite{liu2025deepseek} and the close-source GPT-5 mini~\cite{singh2025openai}. All generation runs use deterministic decoding with temperature set to 0. 
For \approach, we repeat each experimental setting with three independent runs and report the averaged results.

For the structural evolution thresholds introduced in Section~\ref{sec:methodology}, we set the split cohesion threshold to $\gamma_{\mathrm{split}}=2/3$ and the merge coupling threshold to $\theta_{\mathrm{merge}}=0.7$. 
These values are empirical settings selected on Commit0 Lite~\cite{zhao2024commit0} using two held-out repositories, \textit{wcwidth} (small-scale) and \textit{sphinx} (large-scale), to cover repositories of different complexity.  We compared the evolved component-level DAGs against the corresponding golden repository architectures and chose the setting that produced component boundaries closest to the golden architecture by manually comparing the evolved architectures with the corresponding ground-truth repository architectures while avoiding excessive splitting or merging. The held-out repositories are not used in the RepoCraft evaluation.

To isolate the effect of repository-structure planning, each method uses its own method-specific procedure to construct the repository structure. Once the repository architecture is produced, all methods share the same downstream code-generation, validation, and repair scaffold under an identical test-driven development (TDD) execution protocol~\cite{chang2026test,beck2003test}. 

All evaluation-pipeline hyperparameters follow the default RPG configuration~\cite{luo2025rpg}.

Due to space constraints, the main experiment reports three representative RepoCraft repositories: \textit{requests}, \textit{statsmodels}, and \textit{django}, which cover lightweight, medium-scale, and large-scale software projects respectively. All prompts and the results on the remaining three repositories are provided in the supplementary material\supmatfootnote.


\section{Results}

\subsection{RQ1: The effectiveness of \approach}

Table~\ref{tab:rq1-main} presents the main RQ1 results on three representative RepoCraft repositories. Across both backbone models and repositories of different scales, \approach consistently achieves the best overall repository-generation performance.

Under GPT-5 mini, \approach attains the highest \textit{Functionality Coverage} on all three repositories, reaching 100.00\% on \textit{requests}, 80.68\% on \textit{statsmodels}, and 80.50\% on \textit{django}. The same pattern appears under DeepSeek V3.2, where \approach again achieves the best coverage across all evaluated repositories. This indicates that continuous structural evolution helps realize a broader portion of the required functionality.

Beyond covering more required functionality, \approach also translates these improvements into stronger implementation correctness. \approach also achieves the strongest implementation accuracy in most settings. Under GPT-5 mini, it obtains the highest \textit{Pass Rate} on all three repositories, outperforming RPG by +19.47 points on \textit{requests}, +7.61 points on \textit{statsmodels}, and +27.03 points on \textit{django}. For \textit{Voting Rate}, \approach achieves the highest results in all six settings.

The baselines exhibit complementary weaknesses. mini-SWE-agent struggles to maintain repository-wide consistency on larger repositories, especially in \textit{Pass Rate}. Paper2Code often produces relatively high \textit{Functionality Novelty}, but these gains do not consistently translate into stronger correctness. RPG remains the strongest baseline overall, confirming the value of explicit repository-level planning, but it still degrades noticeably on more complex repositories.

Since all methods share the same downstream code-generation, validation, and repair pipeline, the primary difference lies in repository architecture. The consistent improvements therefore indicate that continuously refining repository architecture, rather than relying on a one-shot architectural plan, leads to both broader functionality realization and higher implementation correctness.

\begin{findingbox}{Finding for RQ1}
  \approach consistently achieves the strongest overall repository-generation performance across all baselines by a significant margin.
\end{findingbox}

\subsection{RQ2: Ablation Study}


Table~\ref{tab:rq2-ablation} reports the ablation results of \approach. Removing any individual component degrades performance on at least one repository or evaluation metric, indicating that all four design choices contribute to the final repository-generation quality.

Among all ablations, removing Structural Evolution results in the largest overall performance degradation. Across all three repositories, it consistently reduces both functional coverage and implementation correctness, with particularly large drops on \textit{requests} (\textit{Coverage}: $-5.68$, \textit{Pass Rate}: $-8.47$, \textit{Voting Rate}: $-17.86$) and \textit{django} (\textit{Coverage}: $-5.92$, \textit{Pass Rate}: $-13.33$, \textit{Voting Rate}: $-8.34$). This suggests that repository generation benefits substantially from continuously evolving the architectural state, rather than treating the initially constructed architecture as fixed.

Removing Requirement Context primarily affects implementation correctness. During code generation, the full system conditions each component not only on its directly aligned sub-requirements, but also on the requirement nodes that are functionally coordinated with them through edges in the requirement-level DAG. This additional context helps the model preserve behavior consistency across related functionalities. Removing this contextual information consistently reduces \textit{Pass Rate}, particularly on \textit{django} ($-10.00$) and \textit{requests} ($-5.59$), indicating that neighboring requirement-level context is important for producing implementations that correctly satisfy interacting functional behaviors.

The Component-Graph Ordering mainly influences implementation correctness rather than functional completeness. While \textit{Functionality Coverage} remains unchanged on both \textit{requests} and \textit{statsmodels}, the \textit{Pass Rate} on \textit{statsmodels} decreases by $30.00$ points without dependency-aware generation order, demonstrating that generation order becomes increasingly important as repository complexity grows.

The Dual-DAG representation provides consistent improvements across repositories by separating requirement-level functional coordination from component-level implementation dependencies. Replacing the two-graph representation with a unified graph consistently reduces either functional coverage or implementation accuracy, suggesting that disentangling functional and implementation structures leads to more effective repository planning.

A secondary observation is that several ablated variants achieve slightly higher \textit{Functionality Novelty} on \textit{statsmodels}. Since this metric measures additional generated functionalities rather than successful realization of the benchmark functionality, the increased novelty together with reduced coverage suggests a tendency to generate extra functionalities at the expense of faithfully implementing the required repository behavior.

\begin{table}[t]
\caption{RQ2 ablation results on \textit{requests}, \textit{statsmodels}, and \textit{django}.}
\label{tab:rq2-ablation}
\centering
\scriptsize
\setlength{\tabcolsep}{1.5pt}
\resizebox{\columnwidth}{!}{%
\begin{tabular}{llccc}
\toprule
Repository & Setting & Cov. (\%) & Nov. (\%) & Pass./Vot. (\%) \\
\midrule
\multirow{5}{*}{\textit{requests}}
& \approach & 100.00 & 18.20 & 50.98 / 100.00 \\
& w/o Requirement Context
& 95.45 \abldown{-4.55}
& 9.05 \abldown{-9.15}
& 45.39 \abldown{-5.59} / 87.86 \abldown{-12.14} \\
& w/o Component-Graph Ordering
& 100.00 \ablflat{0.00}
& 12.85 \abldown{-5.35}
& 50.98 \ablflat{0.00} / 100.00 \ablflat{0.00} \\
& w/o Dual-DAG
& 95.45 \abldown{-4.55}
& 9.14 \abldown{-9.06}
& 48.72 \abldown{-2.26} / 87.86 \abldown{-12.14} \\
& w/o Structural Evolution
& 94.32 \abldown{-5.68}
& 8.94 \abldown{-9.26}
& 42.51 \abldown{-8.47} / 82.14 \abldown{-17.86} \\
\midrule

\multirow{5}{*}{\textit{statsmodels}}
& \approach
& 80.68
& 11.48
& 85.51 / 98.65 \\
& w/o Requirement Context
& 67.92 \abldown{-12.76}
& 11.69 \ablup{+0.21}
& 85.51 \ablflat{0.00} / 98.65 \ablflat{0.00} \\
& w/o Component-Graph Ordering
& 80.68 \ablflat{0.00}
& 15.57 \ablup{+4.09}
& 55.51 \abldown{-30.00} / 88.65 \abldown{-10.00} \\
& w/o Dual-DAG
& 78.55 \abldown{-2.13}
& 14.66 \ablup{+3.18}
& 85.51 \ablflat{0.00} / 95.32 \abldown{-3.33} \\
& w/o Structural Evolution
& 75.35 \abldown{-5.33}
& 13.50 \ablup{+2.02}
& 73.51 \abldown{-12.00} / 93.65 \abldown{-5.00} \\
\midrule

\multirow{5}{*}{\textit{django}}
& \approach
& 87.50
& 13.59
& 74.36 / 100.00 \\
& w/o Requirement Context
& 78.29 \abldown{-9.21}
& 12.10 \abldown{-1.49}
& 64.36 \abldown{-10.00} / 100.00 \ablflat{0.00} \\
& w/o Component-Graph Ordering
& 83.56 \abldown{-3.94}
& 11.58 \abldown{-2.01}
& 67.70 \abldown{-6.66} / 100.00 \ablflat{0.00} \\
& w/o Dual-DAG
& 82.24 \abldown{-5.26}
& 12.40 \abldown{-1.19}
& 64.36 \abldown{-10.00} / 93.33 \abldown{-6.67} \\
& w/o Structural Evolution
& 81.58 \abldown{-5.92}
& 10.94 \abldown{-2.65}
& 61.03 \abldown{-13.33} / 91.66 \abldown{-8.34} \\
\bottomrule
\end{tabular}
}
\vspace{-10pt}
\end{table}

\begin{findingbox}{Finding for RQ2}
 All four design components positively contribute to the performance of \approach, while Structural Evolution yields the largest overall improvements.
\end{findingbox}

\subsection{RQ3: Structural Convergence Analysis}

\begin{figure}[t]
  \centering
  \includegraphics[width=\columnwidth]{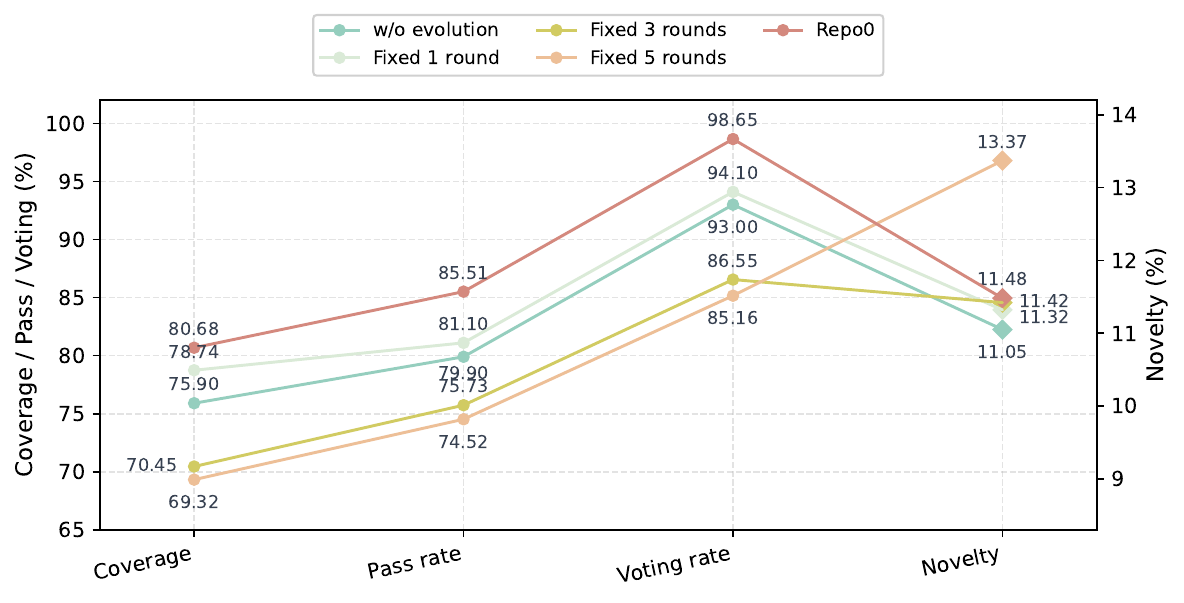}
  \caption{RQ3 structural-convergence analysis on \textit{statsmodels} with GPT-5 mini.}
  \label{fig:rq3-evolution-statsmodels}
\end{figure}

\paragraph{\textbf{Metrics-Guided Convergence}}
Figure~\ref{fig:rq3-evolution-statsmodels} compares \approach with two alternatives on \textit{statsmodels}: (i) \textit{w/o evolution}, which directly uses the initial architectural state, and (ii) LLM-decided structural evolution with fixed budgets of 1, 3, and 5 rounds. Unlike \approach, these budgeted variants allow the LLM to repeatedly perform structural actions without using the proposed cohesion and coupling metrics to determine which components should evolve or when structural convergence has been reached. In contrast, \approach guides \texttt{split} and \texttt{merge} using these metrics and terminates refinement once no metric-triggered structural update remains.

Overall, the full \approach setting achieves the best performance. Compared with \textit{w/o evolution}, \approach improves all four metrics, increasing Functionality Coverage from 75.90\% to 80.68\%, Functionality Novelty from 11.05\% to 11.48\%, Pass Rate from 81.90\% to 85.51\%, and Voting Rate from 93.00\% to 98.65\%. More importantly, \approach also outperforms all budgeted LLM-decided variants, even though those variants are allowed additional structural action rounds. This suggests that the benefit does not simply come from applying more refinement steps; rather, the cohesion and coupling metrics help the component-level DAG reach structural convergence before code generation.

This behavior is consistent with the design of \approach. The two modularity metrics, together with the split and merge decision rules defined in Section~\ref{sec:methodology}, provide an explicit convergence criterion for the evolving component-level DAG. As refinement proceeds, components with diffuse requirement responsibility are split, components with overlapping responsibility are merged, and the loop stops once no further metric-triggered restructuring is needed. This enables subsequent code-generation decisions to operate on a progressively more stable architectural representation.

In contrast, LLM-decided structural evolution lacks this explicit convergence criterion. Repeated structural actions can continue beyond the point of structural convergence, leading to unnecessary fragmentation and compensatory restructuring. Although additional LLM-decided refinement may slightly increase novelty, it degrades coverage, pass rate, and voting rate after one round, indicating that unconstrained structural actions can move the repository away from coherent modular boundaries.

\begin{figure}[t]
\centering
\includegraphics[width=\columnwidth]{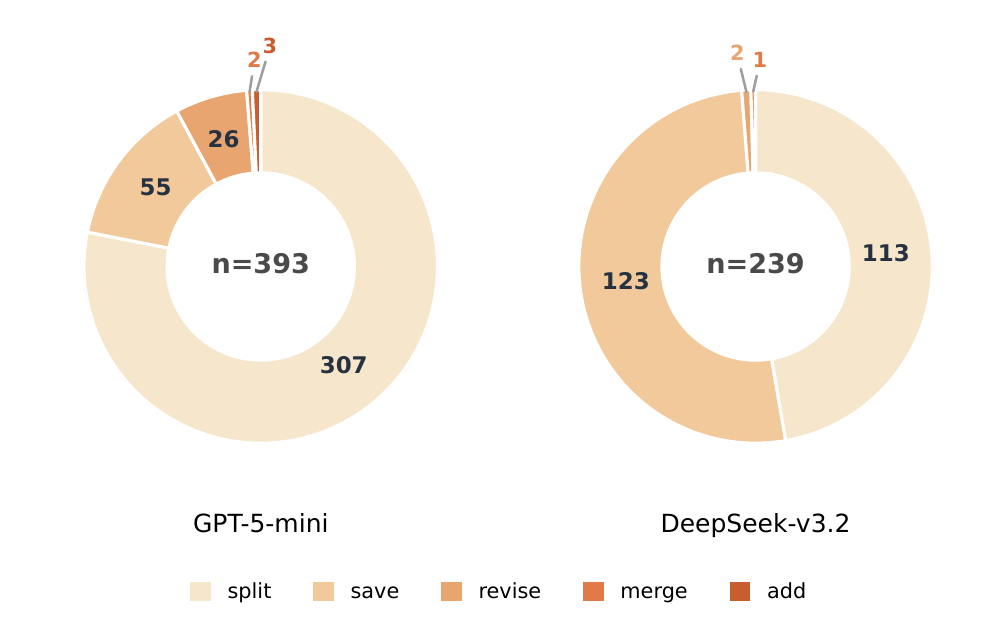}
\caption{Action distributions of different models across the six RepoCraft repositories during structural evolution.}
\label{fig:rq3-action-distribution}
\end{figure}

\paragraph{\textbf{Action Distribution}}
Figure~\ref{fig:rq3-action-distribution} summarizes the aggregated structural-action distribution during metrics-guided structural evolution across both backbone models. In both settings, \texttt{split} is the dominant action, followed by \texttt{save}, while \texttt{merge}, \texttt{revise}, and \texttt{add} occur less frequently. This indicates that structural evolution is primarily driven by localized updates to component boundaries rather than global changes to the component-level DAG. The relatively high proportion of \texttt{save} further suggests that many initial architectural states already contain stable components and require only limited modification.

We further observe systematic differences between backbone models, particularly in the frequency of \texttt{revise} and \texttt{add}. GPT-5 mini triggers these updates more often, and manual inspection confirms that they correspond to valid updates to the architectural state: \texttt{revise} adjusts underspecified component responsibilities and interface assumptions, while \texttt{add} recovers missing requirements or components to improve requirement coverage. In contrast, DeepSeek V3.2 produces fewer such updates, which is consistent with a more complete initial architectural state before structural evolution.

To verify whether this difference is driven by the initial architectural state or the backbone model used during structural evolution, we fix the same unoptimized first-round architectural state generated by GPT-5 mini and apply structural evolution using both backbone models. Under this controlled setting, DeepSeek V3.2 produces nearly the same number of \texttt{revise} actions as GPT-5 mini does when evolving the GPT-5-mini-generated initial architectural state (see the supplementary material\supmatfootnote). This suggests that the action distribution is largely determined by the initial architectural state generated by the LLM. \texttt{revise} actions are therefore especially important for LLMs that produce vague or underspecified initial component responsibilities and interface assumptions.

\begin{findingbox}{Finding for RQ3}
  Metrics-guided convergence improves repository quality beyond unconstrained LLM-decided structural actions, indicating that explicit modularity metrics help the architecture reach structural convergence.
\end{findingbox}

\begin{figure*}[!t]
    \centering
    \includegraphics[width=0.88\textwidth,height=0.52\textheight,keepaspectratio]{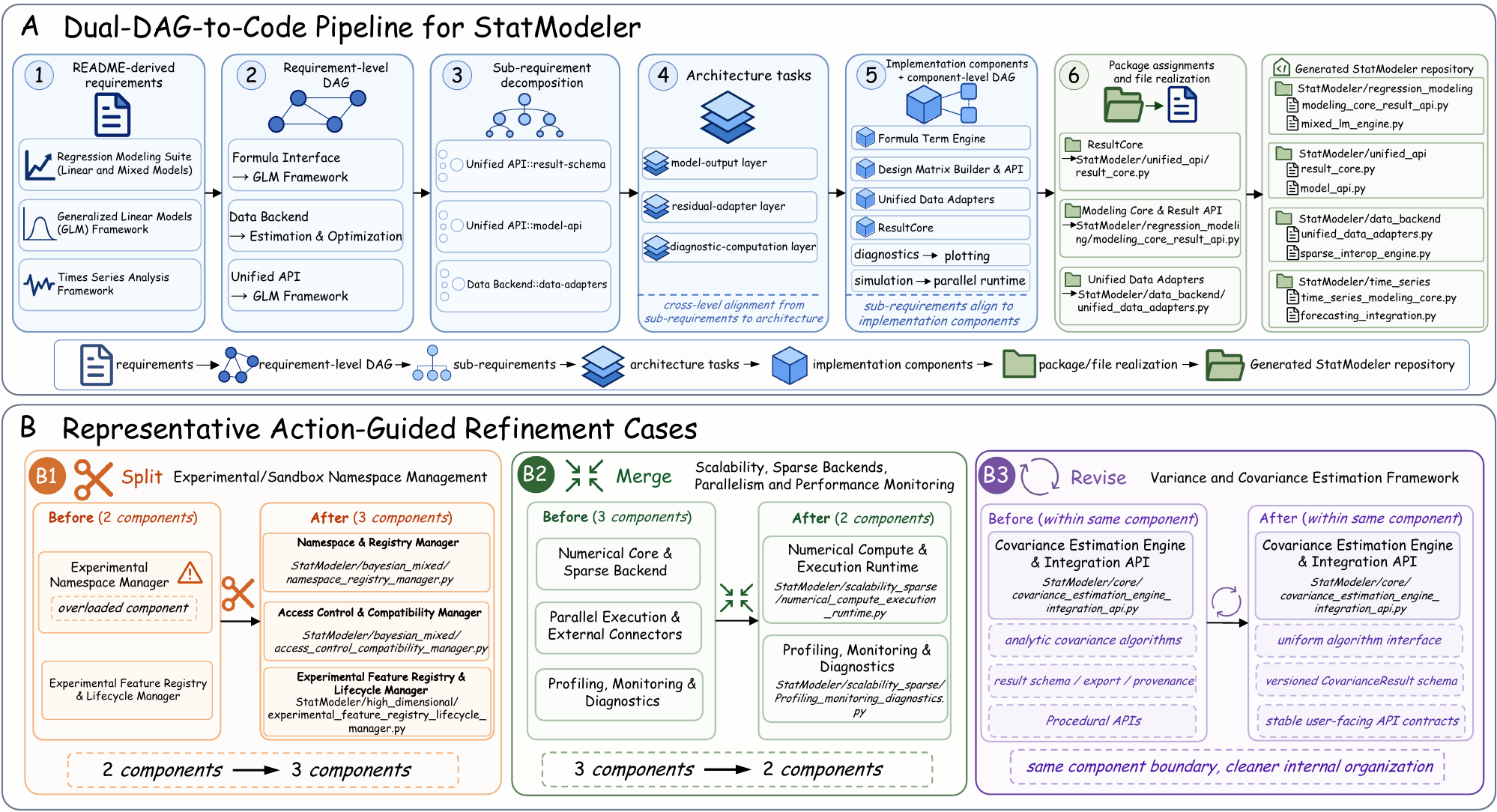}
    \caption{Case study of \approach on StatModeler. The figure shows how requirements are decomposed, aligned with components, updated through structural actions, and materialized into files.}
    \label{fig:case-study}
\end{figure*}

\subsection{Case Study}
Figure~\ref{fig:case-study} shows how \approach instantiates its Dual-DAG on StatModeler, the RepoCraft counterpart of statsmodels. Starting from a README that specifies capabilities such as regression modeling, generalized linear models, time-series analysis, statistical testing, data backends, result serialization, and model diagnostics, the system first extracts 39 high-level requirements as top-level decomposition anchors. These anchors are decomposed and organized into a requirement-level DAG with 92 sub-requirements, such as Unified API::result-schema, Unified API::model-api, and Data Backend::data-adapters. On the implementation side, these sub-requirements are grounded into 86 implementation components, thereby forming the component-level DAG of the Dual-DAG. Throughout this process, the alignment relation explicitly connects sub-requirements to the components that realize them. Components such as ResultCore, Modeling Core \& Result API, and Unified Data Adapters are ultimately materialized as concrete files including StatModeler/unified\_api/result\_core.py, StatModeler/regression\_modeling/modeling\_core\_result\_api.py, and StatModeler/data\_backend/unified\_data\_adapters.py, illustrating how requirement-level structure is carried through to package and file realization.

The figure further highlights how the Dual-DAG remains revisable through structural actions during repository generation. Three representative structural-action cases are shown. A split action refines Experimental Namespace Manager into Namespace \& Registry Manager and Access Control \& Compatibility Manager, making the component boundary more explicit. A merge action consolidates Numerical Core \& Sparse Backend and Parallel Execution \& External Connectors into Numerical Compute \& Execution Runtime, simplifying an over-segmented local structure in the component-level DAG. A revise action keeps the boundary of Covariance Estimation Engine \& Integration API fixed while reorganizing its internals into a cleaner algorithm interface, a versioned CovarianceResult schema, and more stable user-facing API contracts. Together, these cases show that \approach does not treat the Dual-DAG as a fixed blueprint: both component boundaries and the alignment relation can be updated through structural evolution before final code realization.

\section{Cost Analysis}
We analyze the monetary cost of repository generation on the three repositories reported in RQ1, namely \textit{requests}, \textit{statsmodels}, and \textit{django}. Across both DeepSeek V3.2 and GPT-5 mini settings, we decompose costs into generation and evaluation phases to isolate where computational overhead arises.

Under DeepSeek V3.2, \approach incurs generation costs of \$11.95, \$28.19,
and \$27.24 on \textit{requests}, \textit{statsmodels}, and \textit{django},
respectively, while evaluation costs dominate for larger repositories,
particularly \textit{django} (\$72.82), yielding total costs of \$21.28,
\$41.12, and \$100.06. In comparison, RPG exhibits substantially higher
generation overhead on \textit{requests} and \textit{statsmodels} (+\$9.32 and
+\$35.59, respectively), but lower cost on \textit{django} (-\$8.83),
indicating that \approach maintains lower end-to-end cost in most reported
settings while improving repository-generation quality.

Under GPT-5 mini, \approach achieves consistently lower generation cost across all repositories, while evaluation costs remain comparable across methods. The complete cost details for all six repositories are provided in the supplementary material\supmatfootnote. Further analysis shows that improving cohesion and reducing coupling lowers cost more directly by reducing the amount of redundant or conflicting code that needs to be generated and repaired. As a result, \approach requires fewer costly TDD repair iterations caused by ambiguous or overlapping component responsibilities across both backbone models.
%


\section{Threats to Validity}

\textbf{Internal Validity.}
A potential threat is that the quality of structural updates depends on the architectural reasoning capability of the backbone LLM. Although the modularity metrics select candidate \texttt{split} and \texttt{merge} actions, the LLM still instantiates these actions over the current architectural state by rewriting component boundaries, responsibility descriptions, alignment entries, and interface assumptions. This threat is partially mitigated by the iterative structural evolution loop: later evolution rounds can re-evaluate the component-level DAG and the alignment relation under the same metrics, while boundary-preserving \texttt{revise} actions can update component descriptions or interface assumptions without changing component boundaries. During TDD-based code generation, validation feedback can further expose inaccurate component descriptions or interface assumptions, triggering localized repair within the fixed repository architecture.

\textbf{External Validity.}
Our experiments are conducted on six Python repositories from RepoCraft, covering diverse domains and repository scales, but they remain limited to a single programming language and benchmark. The effectiveness of structural evolution on other language ecosystems remains to be investigated. Nevertheless, the proposed architectural state, component-level structural actions, and modularity metrics operate at the level of repository architecture rather than language-specific syntax, suggesting that the framework is not inherently restricted to Python. Evaluating this generality on additional languages and benchmarks is left to future work.

\section{Related Work}

\subsection{Benchmarks for Repository Generation}
Repository generation~\cite{hu2026repo2run,orlanski2026slopcodebench,wang2025codeflowbench,dilgren2025secrepobench,abedu2025repochat,jiang2026doc2featbenchevaluatingdocumentationdrivenfeature,10.1145/3820059,10.1145/3650212.3680347} has recently emerged as a distinct benchmark setting beyond function-level code synthesis. DevBench evaluates broader software-engineering workflows, including implementation, testing, and debugging, but does not primarily target zero-to-repository generation~\cite{chang2026test,li2024devbench,wang2026context}. Commit0 and NL2Repo-Bench move closer to repository-scale tasks, yet both still provide clear structural priors: Commit0 asks agents to complete missing code within a predefined repository architecture, file layout, and function interfaces~\cite{zhao2024commit0}, while NL2Repo-Bench includes the golden repository architecture and detailed module organization in its input specifications~\cite{ding2025nl2repo}.
RepoCraft, introduced together with RPG, tightens the setting toward complete repository construction from high-level natural language requirements while reducing direct exposure to repository-specific implementation structure~\cite{luo2025rpg,zhang2016empirical}. RepoGenesis extends this line to multilingual microservice systems~\cite{peng2026repogenesis}, and ProjDevBench studies end-to-end project development by modern coding agents~\cite{lu2026projdevbench}. These benchmarks progressively reduce structural priors available to repository-generation agents, highlighting the need for methods that can organize and refine repository architectures directly from high-level requirements.
Our work contributes to this line by using modularity principles to drive iterative optimization of repository architecture during generation.

\subsection{Agents for Repository Generation}
Repository-generation agents~\cite{gu2025effectiveness,zhang2026swe,fan2025exploring,wen2018context,zhang2023repocoder,zan2026codes,hu2026line,zeng2025pruning,zeng2026dockerless,zeng2026glimprouter,gao2026swe,huang2027planning,lin2026knowfixqadrivenrepository,chen2026skillforgeselfdistillingagentsprojectspecific,10.1109/ASE63991.2025.00020,shi2025codecorrectnessclosingmile,liu2026effectiveefficientcontextretrieval,rabbi2026multilanguageperspectiverobustnessllm,lin2026reporescueempiricalstudyllm,li2025impactdrivencontextfilteringcrossfile,qin2018syneva} have evolved from role-based workflows toward increasingly explicit structural reasoning. Early systems such as ChatDev, MetaGPT, and SoA coordinate specialized agents across staged development phases, but still rely primarily on natural language or predefined workflows as the carrier of intermediate planning~\cite{chen2025swe,lin2024llms,qian2024chatdev,hong2024metagpt,ishibashi2024self,pan2026persistentcrossattemptstateoptimization,ma2026llmagentscoderepositories}.
Later work introduces more structured planning representations. CodeS decomposes repository generation into repository-, file-, and function-level sketches, enabling hierarchical repository construction through progressively refined planning artifacts~\cite{zan2026codes}. Similarly, Paper2Code employs a multi-stage planning process that derives architectural designs and dependency structures before synthesizing runnable repositories from research papers~\cite{seo2025paper2code}. EvoMAC further improves adaptability by dynamically evolving the multi-agent collaboration topology according to environmental feedback~\cite{hu2025self}. Despite these advances, repository architecture is still treated largely as a planning artifact that is generated once and subsequently executed.

RPG is the closest prior work to our setting. It introduces a Repository Planning Graph that explicitly represents capabilities, file structures, data flows, and functions, moving repository generation from purely natural-language planning toward graph-guided planning~\cite{luo2025rpg}. The key distinction is that RPG treats the planning graph as a blueprint for subsequent generation, whereas \approach treats repository architecture as a persistent and evolving state. Guided by modularity metrics, \approach continuously updates component boundaries and alignments toward higher cohesion and lower coupling.

\section{Conclusion}

This paper presents \approach, a continuous structural evolution framework for repository generation from high-level natural-language requirements. Unlike prior methods that rely on a fixed repository plan, \approach maintains an explicit architectural state and iteratively refines repository structure to improve modularity throughout generation.
Experiments on RepoCraft demonstrate that \approach consistently outperforms representative baselines across both open-source and closed-source backbone models. Ablation studies further show that each major component contributes to overall performance, while metrics-guided structural convergence is critical for effective repository evolution.
Overall, our findings suggest that successful repository generation requires not only code synthesis, but also explicit architectural reasoning and continuous structural refinement. 
We hope this work encourages future research on architecture-aware agents for long-horizon software engineering tasks.


\bibliographystyle{IEEEtran}
\bibliography{ref}

\end{document}